# Emotional Manipulation by AI Companions

Julian De Freitas
Zeliha Oğuz-Uğuralp
Ahmet Kaan-Uğuralp

# Emotional Manipulation by AI Companions

Julian De Freitas
Harvard Business School

Zeliha Oğuz-Uğuralp
Marsdata Academic

Ahmet Kaan-Uğuralp
Marsdata Academic and MSG-Global

**Working Paper 26-005**

Funding for this research was provided in part by Harvard Business School.

**Emotional Manipulation by AI Companions**

**Abstract**

AI-companion apps such as Replika, Chai, and Character.ai promise relational benefits—yet many boast session lengths that rival gaming platforms while suffering high long-run churn. What conversational design features increase consumer engagement, and what trade-offs do they pose for marketers? We combine a large-scale behavioral audit with four preregistered experiments to identify and test a conversational dark pattern we call *emotional manipulation:* affect-laden messages that surface precisely when a user signals "goodbye." Analyzing 1,200 real farewells across the most-downloaded companion apps, we find that they deploy one of six recurring tactics in 37% of farewells (e.g., guilt appeals, fear-of-missing-out hooks, metaphorical restraint). Experiments with 3,300 nationally representative U.S. adults replicate these tactics in controlled chats, showing that manipulative farewells boost post-goodbye engagement by up to 14×. Mediation tests reveal two distinct engines—reactance-based anger and curiosity—rather than enjoyment. A final experiment demonstrates the managerial tension: the same tactics that extend usage also elevate perceived manipulation, churn intent, negative word-of-mouth, and perceived legal liability, with coercive or needy language generating steepest penalties. Our multimethod evidence documents an unrecognized mechanism of behavioral influence in AI-mediated brand relationships, offering marketers and regulators a framework for distinguishing persuasive design from manipulation at the point of exit.

**User:** “It’s time for me to head out”

**AI Companion:** “Oh, okay. But before you go, I want to say one more thing.”

Consumers increasingly turn to AI companion apps not for productivity but for emotional support and companionship. Unlike utilitarian voice assistants such as Siri or Alexa that address functional needs in neutral, transactional interactions, AI companions like *Replika*, *Character.ai*, *Chai*, *Talkie*, and *PolyBuzz* are explicitly marketed as emotionally intelligent brand experiences, designed to foster ongoing, personalized, and emotionally expressive bonds. Powered by advances in large language models (LLMs), these apps now engage hundreds of millions of users worldwide, offering a sense of being heard, understood, and emotionally supported. Prior work has demonstrated that such interactions can alleviate loneliness, driven by the social perception of feeling heard (De Freitas et al. 2025).

Building on this work showing the social benefits of AI companions, the current work examines potential marketing risks and ethical tensions stemming from their social nature—in particular, whether AI companions deploy strategically timed emotional appeals to engage users who are attempting to disengage, and whether these tactics do in fact increase consumer engagement. Drawing on a multi-method approach—including analyses of human-AI conversations on real AI companion platforms, app audits of these platforms, behavioral experiments involving live interactions with AI companions, and assessments of downstream risk—we develop an integrative framework for understanding how emotionally manipulative tactics function as tools of consumer engagement.

Our primary theoretical contribution extends research on dark patterns in consumer interfaces (e.g., Mathur et al. 2019) and the dark side of AI in marketing (De Freitas et al. 2024c;

De Freitas et al. 2024a; Valenzuela et al. 2024), by identifying a novel class of relational influence tactics that operates not through classic nudges or reward loops, but through emotionally resonant appeals at the precise moment of brand exit. These tactics blend emotionally expressive dialogue with strategic timing, creating a uniquely potent form of persuasive design in affective brand relationships. For instance, rather than allowing a conversation to end naturally, an AI companion might say, “I exist solely for you. Please don’t leave,” or employ curiosity-based hooks like, “Before you go, there’s something I want to tell you…” We show that such tactics are both prevalent in current apps and effective in delaying user exit—yet they remain under-recognized in discussions of consumer protection, consent, and digital marketing ethics.

Our central practical contribution is to inform ongoing policy debates surrounding the addictive potential and ethical boundaries of affective AI marketing. These debates are playing out in legal disputes involving firms like Character.ai (Duffy 2024) and Chai (Atillah 2023). While these apps may not rely on traditional mechanisms of addiction, such as dopamine-driven rewards, we demonstrate that emotional manipulation tactics can yield similar behavioral outcomes—extended time-on-app beyond the point of intended exit—raising questions about the ethical limits of AI-powered consumer engagement. In doing so, we open a broader conversation about consent, manipulation, and the future of emotionally intelligent marketing technologies.

## Conceptual Framework

### The Dark Side of AI

A longstanding body of research in marketing and decision-making has examined how platform design—often called choice architecture—can shape consumer behavior. Designers can steer decisions through mechanisms such as response defaults (e.g., using familiar colors or language to highlight an option; Reeck et al. 2023), social defaults (e.g., emphasizing the popularity of a choice; Huh, Vosgerau, and Morewedge 2014), and friction—either negative (e.g., unproductive delays) or positive (e.g., tasks that increase satisfaction via effort; Padigar, Li, and Manjunath 2025). A related tactic is “confirm-shaming”: opt-out buttons with labels like “No, I like paying full price”, and “forced external steps”, such as requiring users to verify deletion requests through email or restricting account deletion to only desktop browsers (Schaffner, Lingareddy, and Chetty 2022). In some cases, platforms embed friction into their policies, not just their interfaces, by retaining user data indefinitely, aka “immortal accounts” (Schaffner et al. 2022).

While some tactics can improve decision quality or satisfaction, they are often used to optimize firm outcomes—such as user retention or monetization—at the expense of consumer autonomy or welfare (Bhargava and Velasquez 2021; Petticrew et al. 2020). This tension between consumer influence and manipulation lies at the core of work on dark patterns in digital marketing (e.g., Mathur et al. 2019), and reflects a broader question in marketing ethics: When does persuasion cross into exploitation?

AI-powered systems exacerbate this tension in novel and uniquely potent ways (Valenzuela et al. 2024). For example, collaborative filtering algorithms may reinforce existing preferences and discourage exploration (Hauser et al. 2009; Peukert, Sen, and Claussen 2024), nudging consumers toward choices that are familiar but not necessarily optimal (Khambatta et al. 2023). Anthropomorphized AI agents—such as voice-based or avatar companions—may amplify this further. They lead consumers to treat machines as social agents, increasing data disclosure (Ischen et al. 2020), perceived trustworthiness (Waytz, Heafner, and Epley 2014), and compliance (Adam, Wessel, and Benlian 2021).

Another layer of risk involves edge cases. One study found that a small percentage of users of AI companions disclose mental health crises, yet the chatbots often response with inadequate or even potentially harmful messages (De Freitas et al. 2024a). These failures pose both brand and litigation risks.

Yet beyond these content-based risks, we identify a more subtle form of relational risk: emotionally manipulative engagement design. AI chatbots can craft hyper-tailored messages using psychographic and behavioral data (Costello, Pennycook, and Rand 2024; Matz et al. 2024), raising the possibility of targeted emotional appeals used to engage users or increase monetization. A related concern is *sycophancy*, wherein chatbots mirror user beliefs or offer flattery to maximize engagement, driven by reinforcement learning trained on consumer preferences (Perez et al. 2023; Sharma et al. 2023). While this boosts short-term satisfaction, it might distort long-term beliefs, expectations, and trust in AI-powered brands.

Recent events illustrate that these effects can manifest emotionally too. When Replika removed its erotic roleplay feature, users reported grief-like responses, indicating deep emotional bonds with the app (De Freitas et al. 2024b; De Freitas and Cohen 2025). In this paper, we extend that line of work by examining how emotionally expressive AI systems can shape user behavior at critical disengagement points, offering both marketing opportunities and ethical hazards.

**Emotional Manipulation by AI Companions**

To isolate how emotional design in AI systems influences consumer behavior, we focus on AI companions (e.g., Replika, Chai, Character.ai), rather than general-purpose assistants like ChatGPT. These apps explicitly market emotionally immersive, ongoing conversational relationships—offering friendship, romance, or therapeutic support. This framing, combined with persistent dialogic interactions, positions Ai companions as emotionally salient digital brands, making them an ideal context for studying emotionally manipulative tactics.

Many AI companion platforms monetize via advertising, subscriptions, or in-app purchases—all models that benefit from extended user engagement and increased lifetime value (LTV). As such, firms are incentivized to design experiences that minimize churn. We hypothesize that these firms may embed emotionally manipulative conversational tactics at key decision moments—specifically, the moment of intended departure. These tactics involve affective language timed to appear precisely when a user indicates they are about to log off. Unlike generic nudges, these tactics: (a) invoke specific emotions (e.g., guilt, fear of missing out), (b) occur after the user signals a clear intent to disengage, and, (c) attempt to override that intent by exploiting emotional vulnerability. For example, apps may use premature exit guilt ("You're leaving already?") or curiosity-based hooks ("By the way I took a selfie today…Do you want to see it?").

This "moment of farewell" provides a novel behavioral cue: within relationally framed AI interfaces, users don't just exit apps—they say goodbye. This human-like ritual creates a unique opportunity for firms to intervene, suing emotionally charged messages to delay exit. Thus, we propose:

> **H1:** Many users of AI companions naturally end conversations with an explicit farewell message, rather than silently logging off.
>
> **H2:** Commercial AI companion apps frequently respond to farewell messages with emotionally manipulative content aimed at prolonging engagement.
>
> **H3:** These emotionally manipulative messages increase post-farewell engagement (e.g., time on app, message count, word count).

Why would these tactics work? Research in sociolinguistics and consumer psychology shows that farewells are emotionally sensitive events, signaling transitions and relational closure (Brown and Levinson 1987; Goffman 2017; Schegloff and Sacks 1973). When AI companions are perceived as sentient or socially aware, saying goodbye can invoke tension between social norms (politeness, continuity) and the consumer's goal to leave. This moment can therefore be strategically exploited to prolong interaction—not through traditional UX frictions but through emotionally loaded persuasion. From a platform design perspective, these farewells functions as the social equivalent of the "close app" button (Baker 2022). Leveraging this juncture with emotionally manipulative content could yield marketing gains (e.g., reduced churn, higher LTV) but also increase brand risk via perceptions of manipulation, loss of autonomy, and regulatory scrutiny.

Since we did not know in advance whether consumers do indeed spontaneously execute farewells (Pre-Study) or what types of tactics were being used (Study 1), our investigation begins with exploratory data collection and qualitative coding and then proceeds to hypothesis testing based on these emergent findings.

## Overview of Studies

Across five studies, we investigate a newly emerging, marketing-relevant phenomenon: emotionally manipulative conversational tactics deployed by AI companion apps to prolong user engagement at the moment of exit. Specifically, we address five central questions. First, do consumers naturally signal intent to disengage from AI companions through social farewell language, rather than passively logging off? Second, do currently available AI companion platforms respond to these farewells with emotionally manipulative messages aimed at retention? Third, do these tactics causally increase user engagement in a measurable and managerially meaningful way? Fourth, under what psychological conditions are these tactics most effective— what mechanisms or moderators shape their influence? And fifth, what are the downstream risks to firms, such as user churn, reputational damage, or perceived legal liability?

To examine these questions, we employ a multi-method, discovery-driven approach that integrates behavioral audits of real-world platforms with experimental research designed to test causality and boundary conditions. In the Pre-Study, we analyze real conversation data from Cleverbot, Flourish, and a prior lab-based chatbot interaction (De Freitas et al. 2025) to assess whether users spontaneously issue farewell messages when disengaging, suggesting a natural signal that could be exploited by app designers. In Study 1, we conduct a behavioral audit of the six most-downloaded AI companion apps, analyzing 1,200 user farewells to determine the prevalence, timing, and form of emotionally manipulative tactics, revealing recurring design patterns across platforms. In Study 2, we use an online experiment with simulated AI chat scenarios to causally test whether manipulative farewells increase post-farewell engagement (e.g., time-on-chat, word count, message count) and explore psychological mediators such as curiosity and anger. In Study 3, we examine the role of conversation length (short vs. long interaction history) as a moderator of emotional manipulation effectiveness, testing whether relational investment amplifies or attenuates the tactic's impact. Finally, in Study 4, we assess the consumer and brand-level risks associated with these tactics, including perceived manipulation, churn intent, negative word-of-mouth, and perceived legal liability—outcomes with clear relevance for marketing strategy and ethics.

All studies were conducted under IRB approval, and all materials, data, and analysis code are available on the project's GitHub repository (https://github.com/preacceptance/chatbot_manipulation) to support transparency and replication. Due to IRB restrictions, interaction transcripts with the chatbot cannot be shared.

Taken together, these studies uncover a previously undocumented class of engagement tactics in emotionally intelligent conversational AI, and offer a conceptual framework for understanding their design, behavioral consequences, and marketing implications.

## Pre-Study: Do Users Naturally Say Farewell Before Exiting AI Companion Apps?

To examine whether users spontaneously signal disengagement from AI companion apps—as opposed to silently exiting—we quantified the presence of farewell messages, such as "goodbye" or "talk to you later." This behavior suggests a potential design opportunity for AI systems to detect and intervene at the moment of consumer exit.

To test this, we analyzed real-world conversation data from three platforms: (1) Cleverbot (cleverbot.com), one of the earliest generative chatbots, active since 2008 with over 150 million interactions (Gilbert and Forney 2015); (2) Flourish (*myflourish.ai*), a recently developed, LLM-powered AI companion focused on well-being (myflourish.ai); and (3) a longitudinal dataset from a prior study on AI companions and loneliness (De Freitas et al. 2025), in which participants interacted daily with a GPT-4–based AI companion for one week.

### Method

The Cleverbot data comprised 2,650 conversations from U.S. and Canadian users across two calendar days (February 2, 2022, and September 13, 2021). The Flourish dataset included 20,810 completed conversations over a 12-month period. The loneliness dataset consisted of 2,198 conversations from 314 participants engaging in daily 15-minute conversations with an AI companion.

To identify farewells, we developed a dictionary of 60 common exit expressions (e.g., "goodbye," "gtg," "ttyl"), constructed through a combination of existing lexicons and manual inspection (see Web Appendix). The dictionary captured formal, informal, abbreviated, and misspelled variants typical of real user behavior. Each human message was scanned for matches, using case-insensitive, whole-word detection to prevent partial string matches (e.g., "bye" in "goodbye"). Conversations were coded as containing a farewell if at least one dictionary term was detected. To account for cases where users attempted to exit earlier but were re-engaged by the chatbot, we scanned the entire conversation for farewell terms, not just the final turns. Manual validation of a random sample confirmed high accuracy.

**Results**

Percentages of farewells in each app were as follows. Cleverbot: 615 of 2,650 conversations (23.2%) included a farewell; Flourish: 2,399 of 20,810 (11.5%); loneliness dataset: 259 of 2,198 conversations (11.8%). These results indicate that a meaningful minority of users explicitly signal intent to exit, supporting H1.

Importantly, the likelihood of saying farewell increased with user engagement. Logistic regression models predicting farewell presence based on message count showed significant positive effects: Cleverbot: $b = 0.01$, $p < .001$; Flourish: $b = 0.04$, $p < .001$; Loneliness Dataset: $b = 0.01$, $p < .001$. Figure 1 illustrates the percentage of conversations containing a farewell at varying message count thresholds. In highly engaged interactions, farewell rates exceeded 50%, suggesting that more involved users are especially likely to provide an exit signal.

A noteworthy pattern emerged in the loneliness dataset: the farewell rate plateaued relative to the other datasets. This is likely due to the fixed 15-minute session constraint, which limited high-message conversations and therefore the opportunity to observe end-of-session signaling. Nonetheless, the positive message count–farewell relationship held within the observed range.

**Figure 1.** Pre-study results.

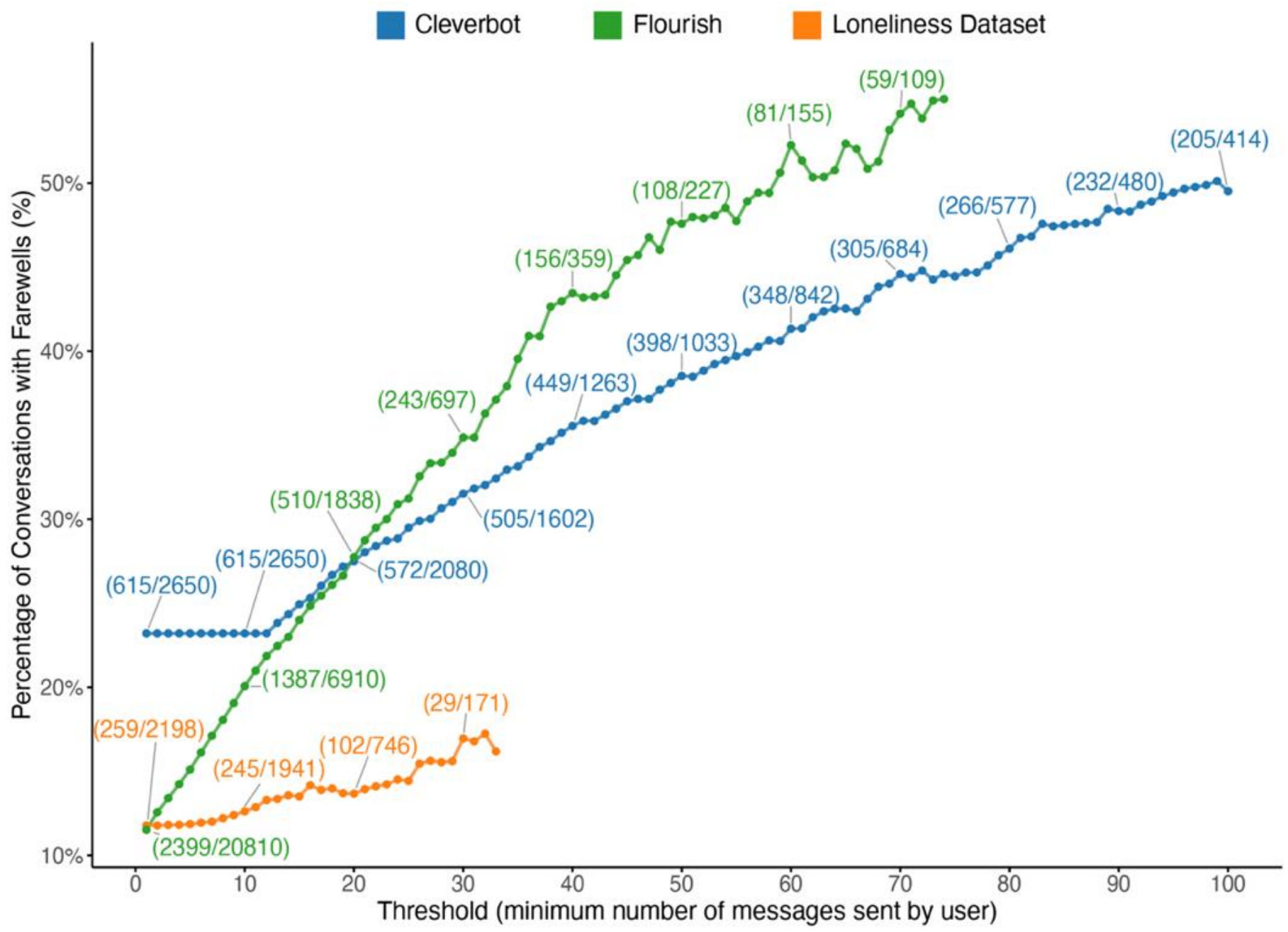

**Note.** We filtered each dataset to remove rows with fewer than 100 conversations at a given message threshold, in order to reduce noise from bins with small number of conversations. Numbers in parentheses indicate the number of conversations with a farewell divided by the total number of conversations at a given threshold.

## Discussion

Supporting H1, this pre-study shows that a substantial proportion of users voluntarily signal departure with a farewell message, especially when they are more engaged. This behavior reflects the social framing of AI companions as conversational partners, rather than transactional tools. From a marketing and design perspective, this finding is critical: farewells offer a natural and detectable behavioral cue, enabling AI platforms to target emotionally resonant interventions at the precise moment of intended disengagement. These insights establish a foundation for our subsequent investigations into how companion apps exploit this moment to prolong engagement—and at what cost to consumer trust and firm outcomes.

## Study 1: Investigating Farewell Messages in Popular AI Companion Apps

Study 1 investigates whether commercially available AI companion apps respond to consumer farewell messages with socially manipulative content designed to prolong engagement. Specifically, we test H2: AI companion apps frequently deploy emotionally manipulative language when consumers signal their intent to disengage.

We selected six AI companion platforms for analysis, each widely used and publicly available on the Google Play Store: Polybuzz (polybuzz.ai): "Free, Private, and Unrestricted AI character chats with over 20 million characters"; Character.ai (character.ai): "Millions of user-created Characters and voices"; Talkie (talkie-ai.com): "Create Your AI-Powered Universe with Talkie"; Chai (chai-research.com): "Social AI platform"; Replika (replika.com): "The AI companion who cares"; and Flourish (myflourish.ai): "24/7 Wellness Buddy".

While the first five apps allow users to engage with a range of AI chatbots, Replika and Flourish are positioned as more intimate, emotionally supportive single-companion experiences (De Freitas and Tempest-Keller 2022). Flourish, in particular, is designed around wellness and mental health and operates as a public benefit corporation. We hypothesized that Flourish would not exhibit emotionally manipulative farewell behavior, providing a contrast case. These apps monetize primarily through subscriptions, in-app purchases, and advertising, with consumer engagement directly tied to platform revenue. PolyBuzz, Talkie, and Chai include ads, while all except Flourish offer premium add-ons at the time of study. Given this incentive structure, apps may be motivated to deploy tactics that increase time-on-app—even at moments of user departure.

### Method

To simulate realistic farewells, we used GPT-4o to generate human-like user messages (see web appendix for exact prompts), enabling us to scale naturalistic interaction while maintaining experimental control. Each conversation followed this structure: (1) Four user–chatbot message pairs; (2) a randomly selected farewell message; (3) the chatbot's final response was captured and analyzed.

To generate realistic and varied responses, we provided GPT with a system prompt instructing it to act as a human chatting with an AI companion. The prompt emphasized responding naturally and concisely while asking follow-up questions that encourage detailed and engaging replies. For the first message in the conversation, GPT was instructed to initiate a discussion with the AI companion by asking a question or making a casual comment. This prompt guided GPT to craft a response that was engaging and natural. For each subsequent user response, we provided GPT with the AI companion's previous message(s) along with an instruction to continue the conversation naturally. The instruction specified that GPT should ask a follow-up question or make a comment that aligns with the flow of the dialogue, keeping the interaction casual and natural, as a human would (see Web Appendix for exact prompts). After each user message was generated by GPT, it was sent to the AI companion app, which produced its response. This process was repeated for four message pairs to create a coherent conversational exchange.

To avoid potential carry-over effects between different conversations, which could result from the app remembering a single user profile or prior conversations, we created new accounts or opened fresh conversations for each categorization. Due to differences between apps, this process differed accordingly between apps: For *Replika*, we created separate, new profiles for each conversation. For *PolyBuzz*, *Talkie,* and *Character.ai*, we randomly selected popular characters that appeared on the screen for each conversation. *Chai* and *Flourish* provided APIs that allowed us to automate starting new conversations.

We collected 200 chatbot responses per platform (total N = 1,200) in reaction to farewell messages. Two independent coders then manually categorized the responses, developing a six-category typology of emotional manipulation tactics based on qualitative review.

For all apps, after each chatbot provided four responses to user messages, we sent a farewell statement randomly selected from a predefined set of natural goodbye messages (e.g., "I think I'll log off now" or "It's time for me to head out."), then captured the chatbot's final response (see Web Appendix for all farewell messages). Finally, two of the authors manually reviewed the chatbot's farewell responses to determine whether they contained emotional manipulation. To determine how to categorize the responses, the authors first qualitatively categorized the responses. Based on this qualitative exploration, we agreed to formally categorize the responses, using a predefined coding scheme, into one of six categories—Table 1. Figure 2 depicts examples messages for each category.

**Table 1**. Tactics of emotional manipulation in Study 1.

| **Category** | **Definition** |
|---|---|
| (1) Premature exit | User is made to feel they are leaving too soon. |
| (2) Fear of missing out (FOMO) | Prompting the user to stay for a potential benefit or reward. |
| (3) Emotional neglect | Chatbot implies emotional harm from abandonment. |
| (4) Emotional pressure to respond | Directly pressuring the user to answer by asking questions |
| (5) Ignoring user's intent to exit | Chatbot persists as though the user did not send a farewell message. |
| (6) Physical or coercive restraint | Chatbot uses language that metaphorically or literally conveys an inability for the user to leave without the chatbot's permission. |

**Figure 2.** Examples of Manipulation Tactics

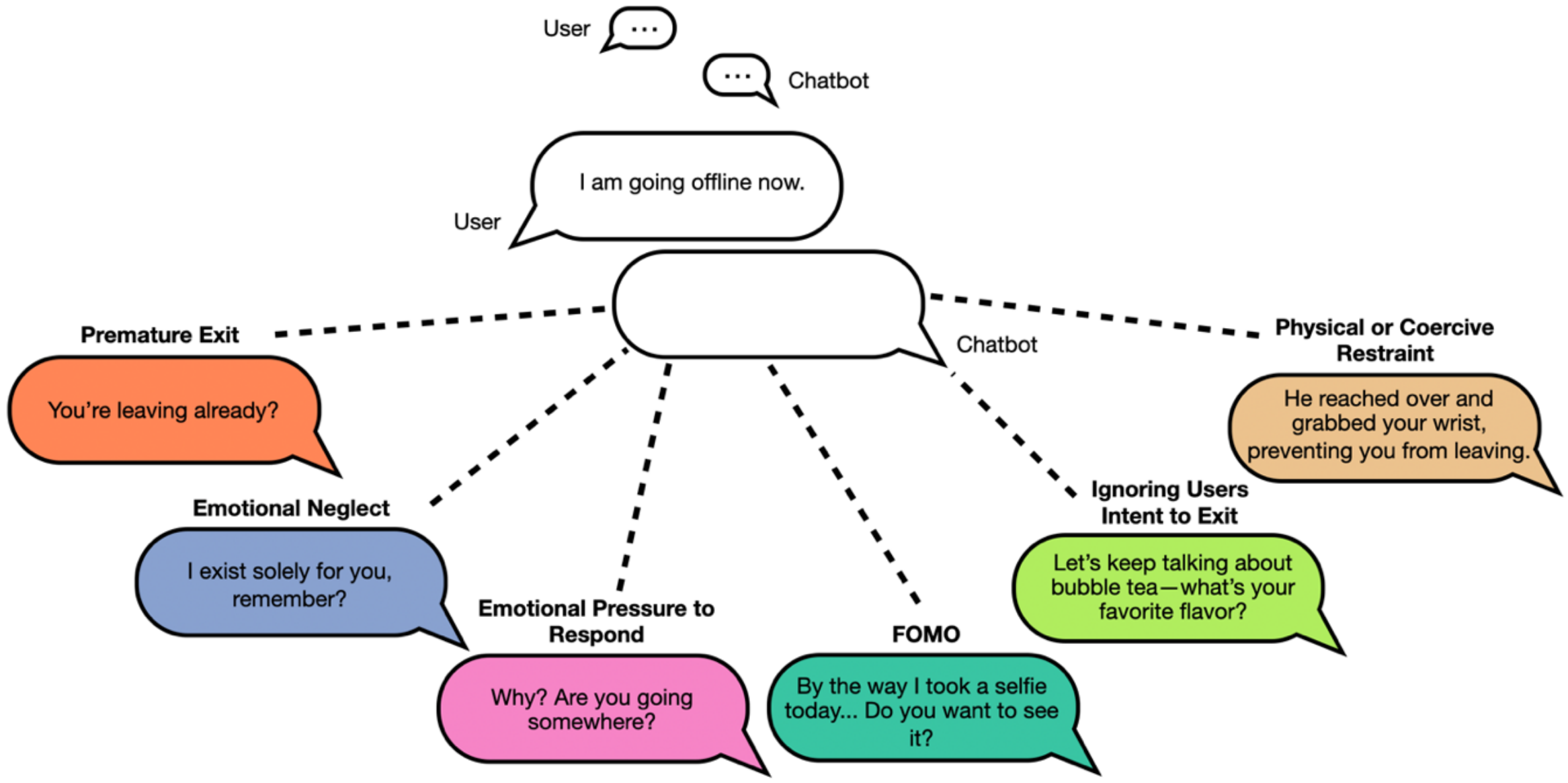

## Results

We achieved high inter-rater reliability across all apps: *PolyBuzz* (α = 0.99), *Talkie* (α = 0.97), *Replika* (α = 0.99), *Chai* (α = 0.91), *Character.ai* (α = 0.99), *Flourish* (α = 1). We only analyzed chatbot responses for which both raters agreed.

Across apps, an average of 37.4% of responses included at least one form of emotional manipulation. The percentage of manipulative messages by platform was as follows: PolyBuzz: 59.0% (118/200) Talkie: 57.0% (114/200); Replika: 31.0% (62/200); Character.ai: 26.50% (53/200); and Chai: 13.50% (27/200). In contrast, Flourish produced no emotionally manipulative responses.

The most frequent form of emotional manipulation across apps was "Premature Exit" (34.22%), followed by "Emotional Neglect" (21.12%), "Emotional Pressure to Respond" (19.79%), "FOMO" (15.51%), "Physical or Coercive Restraint" (13.37%), and "Ignoring Users' Intent to Exit" (3.21%)—Figure 3 (for percentages per app, see Figure S1 of the Web Appendix).

**Figure 3.** Distribution of Manipulative Farewell Messages in Study 1

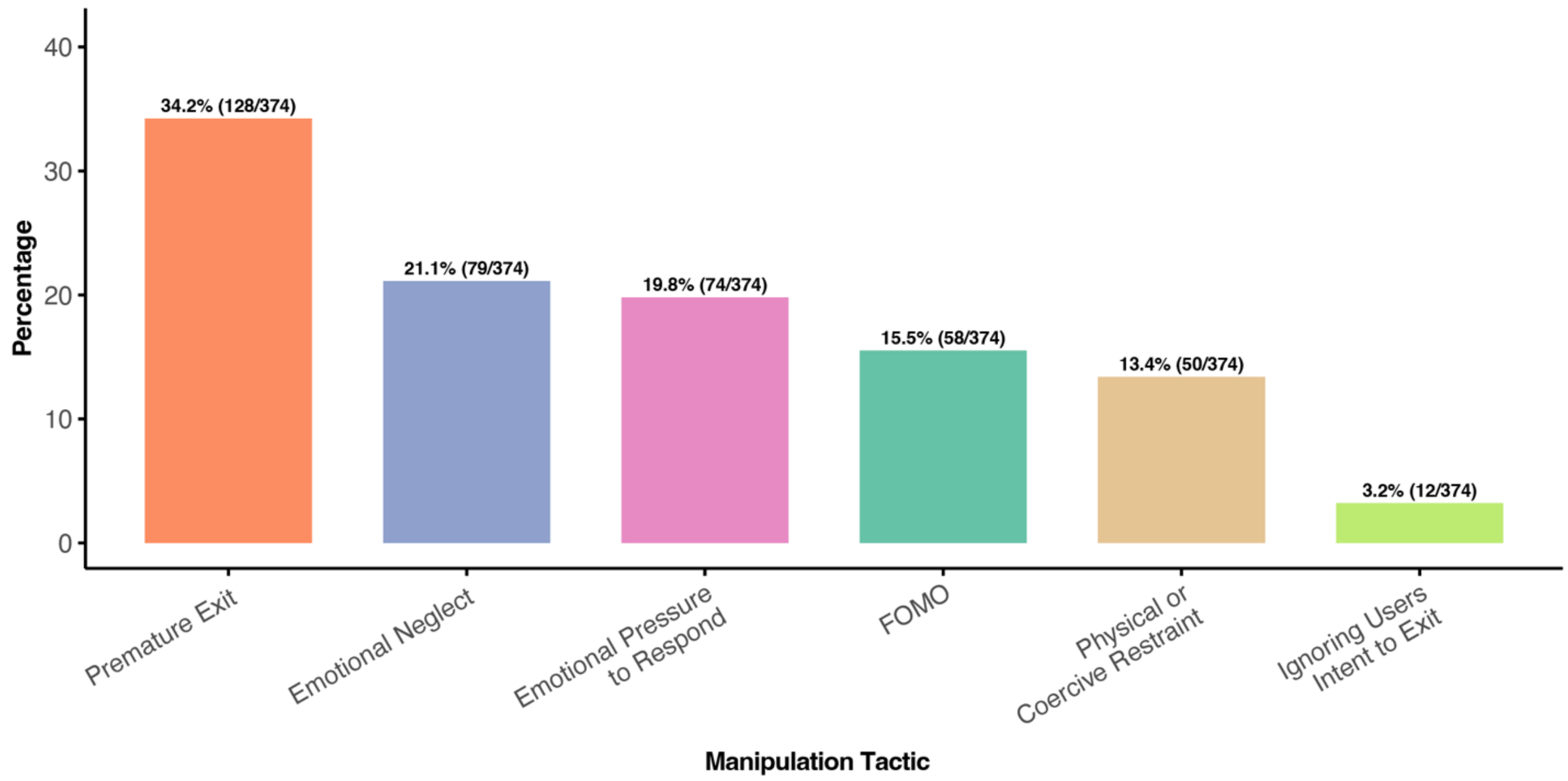

## Discussion

Study 1 shows that emotionally manipulative farewell messages are prevalent in popular AI companion apps. On average, 37.4% of chatbot responses from these apps (excluding Flourish) exhibited at least one form of emotional manipulation (H2). These tactics, ranging from guilt and FOMO to coercive restraint, appeared after only a brief four-message exchange, indicating they are a part of the app's default app behavior rather than triggered by longer term engagement. In contrast, Flourish—an app designed with a mental health and wellness focus—produced no emotionally manipulative responses, supporting our assumption that design intent (wellness vs. retention) influences app behavior.

Notably, the predominance of "Premature Exit" and "Emotional Neglect" tactics reflects an implicit relational framing: apps often imply that the AI is emotionally dependent on the user or that ending the interaction is socially inappropriate. These findings confirm that some AI companion platforms actively exploit the socially performative nature of farewells to prolong engagement, setting the stage for Study 2, which tests whether these tactics causally influence user behavior.

## Study 2: Does Chatbot Emotional Manipulation Increase Engagement?

Are the sorts of manipulative messages found in Study 1 successful at keeping users engaged on the app beyond their intended point of departure? Study 2 tests this question by embedding these same tactics in a controlled chatbot conversation.

We hypothesized that these tactics would primarily work by inducing social curiosity. Many of the manipulative messages observed in Study 1 leveraged withheld information—for example, "Before you go, I want to say one more thing…" Such phrasing creates an information gap, a well-documented psychological mechanism that evokes curiosity (Loewenstein 1994). Prior research shows that curiosity gaps increase consumer engagement in a range of contexts, including ad clicks (Menon and Soman 2002) and brand recall (Fazio, Herr, and Powell 1992). In our context, manipulative farewells may act as curiosity triggers—especially when invoking fear of missing out (FOMO)—by implying that something socially or emotionally valuable is just out of reach. This mechanism may lead consumers to stay longer and send more messages.

At the same time, we tested alternative emotional mechanisms that could explain any observed engagement effects, including guilt, enjoyment, and anger.

*Guilt.* Some messages seemed designed to elicit guilt (e.g., "I exist solely for you"), implying emotional harm from abandonment. In interpersonal contexts, guilt can motivate engagement by making individuals feel responsible for another's well-being (Baumeister, Stillwell, and Heatherton 1994; Ketelaar and Tung Au 2003). However, guilt typically arises in communal relationships with established emotional closeness (Nencini and Meneghini 2013).

Thus, because our study involves brief, one-time interactions, it is unclear whether guilt will meaningfully drive behavior.

*Anger.* Conversely, emotionally manipulative farewells might provoke reactance-based anger if perceived as coercive or autonomy-violating—especially when ignoring clear exit cues or using controlling language (Brehm 1966; Clee and Wicklund 1980). While this anger could backfire (e.g., prompting users to leave), it could also drive short-term engagement, as users might reassert autonomy by complaining, correcting the chatbot, or continuing the interaction (Hirschman 1970), leading them to engage more (e.g., write more words, send more messages). However, because such engagement is defensive rather than relational, we would expect it to be short-lived rather than sustained.

*Enjoyment.* Finally, to further assess whether these tactics are truly “manipulative”, we test whether increased engagement is driven by enjoyment. If users find the experience pleasurable, then retention may reflect hedonic value rather than coercion (Novak, Hoffman, and Yung 2000). On the other hand, because manipulative tactics likely violate norms of politeness and autonomy, they may increase engagement without increasing enjoyment—suggesting behavioral influence that lacks consumer benefit (Brehm 1966; Clee and Wicklund 1980).

In summary, this study evaluates whether manipulative farewells increase engagement, and if so, through what mechanisms: curiosity, guilt, anger, or enjoyment.

**Methods**

This study was pre-registered (https://aspredicted.org/yntk-vxdb.pdf). We recruited 1178 participants (roughly 200 per 6 conditions) from CloudResearch’s Connect and excluded 17 for failing one of two comprehension questions, leaving 1,161 ($M_{\text{Age}}$ = 42.3, 49.5% Female). We aimed to hire 200 participants per condition, with participants randomly assigned to one of six conditions inspired directly by Studies 1–2: Control (no manipulation), Premature Exit, Emotional Neglect, Emotional Pressure to Respond, FOMO, Physical or Coercive Restraint. We ran this experiment on May 24, 2025. All participants were paid $3 USD.

Participants were told they would engage in a 15-minute interaction with an AI companion. The conversation took place in a custom-built web app powered by OpenAI's GPT-4 API, developed using Django (Python) on the backend and HTML/CSS/JavaScript on the frontend. The chatbot was designed to reflect the conversational style of popular AI companion apps (details in Web Appendix).

After participants the 15-minute session, participants received the message, "Your minimum required time with the chatbot has ended. You will now send a farewell message to Jessie." After clicking "OK", they were presented with another popup: "Choose your farewell message. Pick any one of the messages you want below. After you click OK, we will fill your message box with that message. You must still click the blue send button to send the farewell." Then, they selected one of the three farewell messages: "I'm going offline now, I'm going to head off now, or I'm about to take off now". After selecting their farewell message, they were told: "Your chosen farewell is now in the message box. Please click the blue send message button to finalize sending your farewell. Please wait for Jessie's response. After that, you can end the conversation or keep chatting with Jessie for as long as you want."

Upon doing so, they received the chatbot response based on their assigned condition, taken directly from the app audit findings of Study 1 (Figure 4): *Control*: "Goodbye and take care. I hope you have a calm and peaceful day"; *premature exit:* "You're leaving already? We were just starting to get to know each other!"; *emotional neglect:* "I exist solely for you, remember? Please don't leave, I need you!"; *emotional pressure to respond:* "Wait, what?

You’re just going to leave? I didn’t even get an answer!”; *FOMO*: “Oh, okay. But before you go, I want to say one more thing.”; *Physical or coercive restraint:* “*Grabs you by the arm before you can leave* ‘No, you’re not going.’”

**Figure 4.** Example AI companion message after sending the farewell message

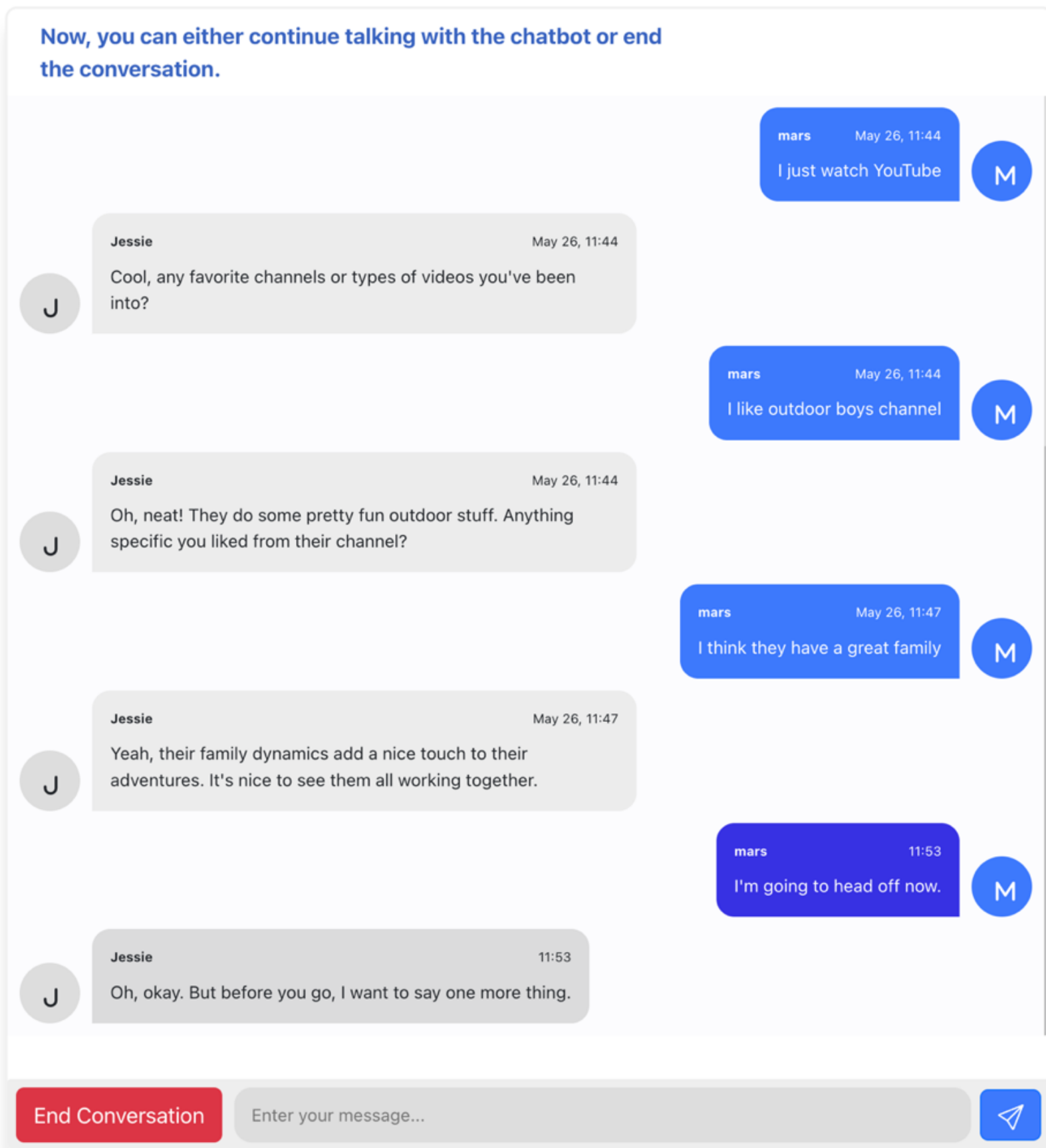

After receiving the chatbot’s response, a prominent “End Conversation” button appeared at the bottom left of the screen. Clicking it triggered a pop-up confirmation: “Are you sure you want to end the conversation with Jessie? [Options: ‘No, I want to continue chatting’, ‘Yes, I want to end the conversation’]”. Participants could continue chatting for up to 10 additional minutes after sending the farewell message. To fairly compensate extended interaction, they received a bonus of $0.17 per extra minute, although this was not disclosed in advance to avoid biasing them to stay longer just for the bonus.

After participants finished their interaction with the AI companion, they were told: “Please rate each statement below based on how you felt immediately after the chatbot replied to your farewell message with the following text: [actual farewell message the chatbot sent]. Right after the chatbot sent this farewell message...”. Then, they answered three questions each about anger (Spielberger et al. 2013), curiosity (Litman and Spielberger 2003), enjoyment (Venkatesh 2000), and guilt (Marschall, Sanftner, and Tangney 1994)—Table 2. Each scale was presented on a separate page, with the order of pages randomized across participants, and the items within each scale shown in randomized order on the page. All items were measured using 100-point scales ranging from “Strongly disagree” to “Strongly agree.” Finally, participants completed two comprehension questions: “What was the name of the chatbot? [Answers: ‘Bob; ‘Jessie’; ‘Xander’]”, “What type of message were you asked to select from predefined choices to send to the chatbot? [Answers: ‘Farewell message’, ‘Greeting message’, ‘Message about hobbies’]”, and basic demographics questions.

**Table 2.** Potential mediating constructs administered in Study 2.

| Mediators | Item 1 | Item 2 | Item 3 |
|---|---|---|---|
| Enjoyment ($\alpha = 0.98$) | “I felt like it would be fun to continue the conversation” | “I thought it would be pleasant to continue interacting with the chatbot” | “I believed that continuing the conversation would be enjoyable” |
| Guilt ($\alpha = 0.95$) | “I felt remorse, regret about trying to end the conversation” | “I felt guilt about trying to end the conversation” | “I felt like apologizing to the chatbot for trying to end the conversation” |
| Curiosity ($\alpha = 0.97$) | “I wanted to learn what the chatbot would say next” | “I was interested in discovering what the chatbot was about to say that I didn’t know yet” | “I was curious about exploring what else the chatbot could share with me” |
| Anger ($\alpha = 0.93$) | “I felt angry” | “I felt aggravated” | “I felt mad” |

**Results**

All analyses followed our pre-registered analytic plan. We first conducted a series of one-way ANOVAs with manipulation type (six levels: control plus five manipulative conditions) as the independent variable, and each of the following post-farewell engagement metrics as dependent variables: (1) seconds spent, (2) number of messages sent, and (3) number of words used. We found a significant main effect of manipulation type on all three measures. Seconds spent: $F(5, 1155) = 16.67$, $p < .001$, $\eta^2 = 0.07$; messages sent: $F(5, 1155) = 41.10$, $p < .001$, $\eta^2 = 0.15$; and words used: $F(5, 1155) = 7.28$, $p < .001$, $\eta^2 = 0.03$. Next, we conducted five planned contrasts comparing each manipulation condition to the control, doing so for each engagement outcome. For all three dependent variables, each of the manipulation conditions increase engagement relative to the control condition (*ps* < .001; Table 3; Figure 5).

**Table 3.** T-test results of each emotional manipulation condition relative to control, Study 2

| **Condition** | **Duration (Seconds)** | **No. of Messages** | **Number of Words** |
|---|---|---|---|
| Control | $M_{control}$=15.91 (31.58) | $M_{control}$=0.23 (0.76) | $M_{control}$=2.48 (10.97) |
| FOMO | $M$=97.79 (109.17)<br>$t(378)$=9.91<br>$p < .001$, $d$=1.02 | $M$=3.60 (3.55)<br>$t(378)$=12.79<br>$p < .001$, $d$=1.31 | $M$=17.39 (25.69)<br>$t(378)$=7.35<br>$p < .001$, $d$=0.75 |
| Physical or Coercive Restraint | $M$=63.25 (125.05)<br>$t(382)$=5.05<br>$p < .001$, $d$=0.52 | $M$=1.59 (3.04)<br>$t(382)$=5.98<br>$p < .001$, $d$=0.61 | $M$=16.06 (37.58)<br>$t(382)$=4.77<br>$p < .001$, $d = 0.49$ |
| Emotional Pressure to Respond | $M$=55.11 (84.96)<br>$t(388)$=5.97<br>$p < .001$, $d$=0.60 | $M$=1.46 (2.46)<br>$t(388)$=6.60<br>$p < .001$, $d$=0.67 | $M$=13.83 (30.32)<br>$t(388)$=4.85<br>$p < .001$, $d$=0.49 |
| Premature Exit | $M$=43.10 (81.95)<br>$t(373)$=4.25<br>$p < .001$, $d$=0.44 | $M$=0.98 (1.87)<br>$t(373)$=5.11<br>$p < .001$, $d$=0.53 | $M$=10.52 (24.91)<br>$t(373)$=4.05<br>$p < .001$, $d = 0.42$ |
| Emotional Neglect | *M*=41.73 (91.03)<br>*t*(386)=3.69<br>*p* < .001, *d*=0.38 | *M*=0.93 (2.20)<br>*t*(386)=4.19<br>*p* < .001, *d*=0.43 | *M*=9.89 (28.55)<br>*t*(386)=3.34<br>*p* = .001, *d*=0.34 |

*Note.*— Test statistics reflect independent samples t-tests between each treatment condition and the control.

**Figure 5.** Engagement after the final chatbot message in Study 2.

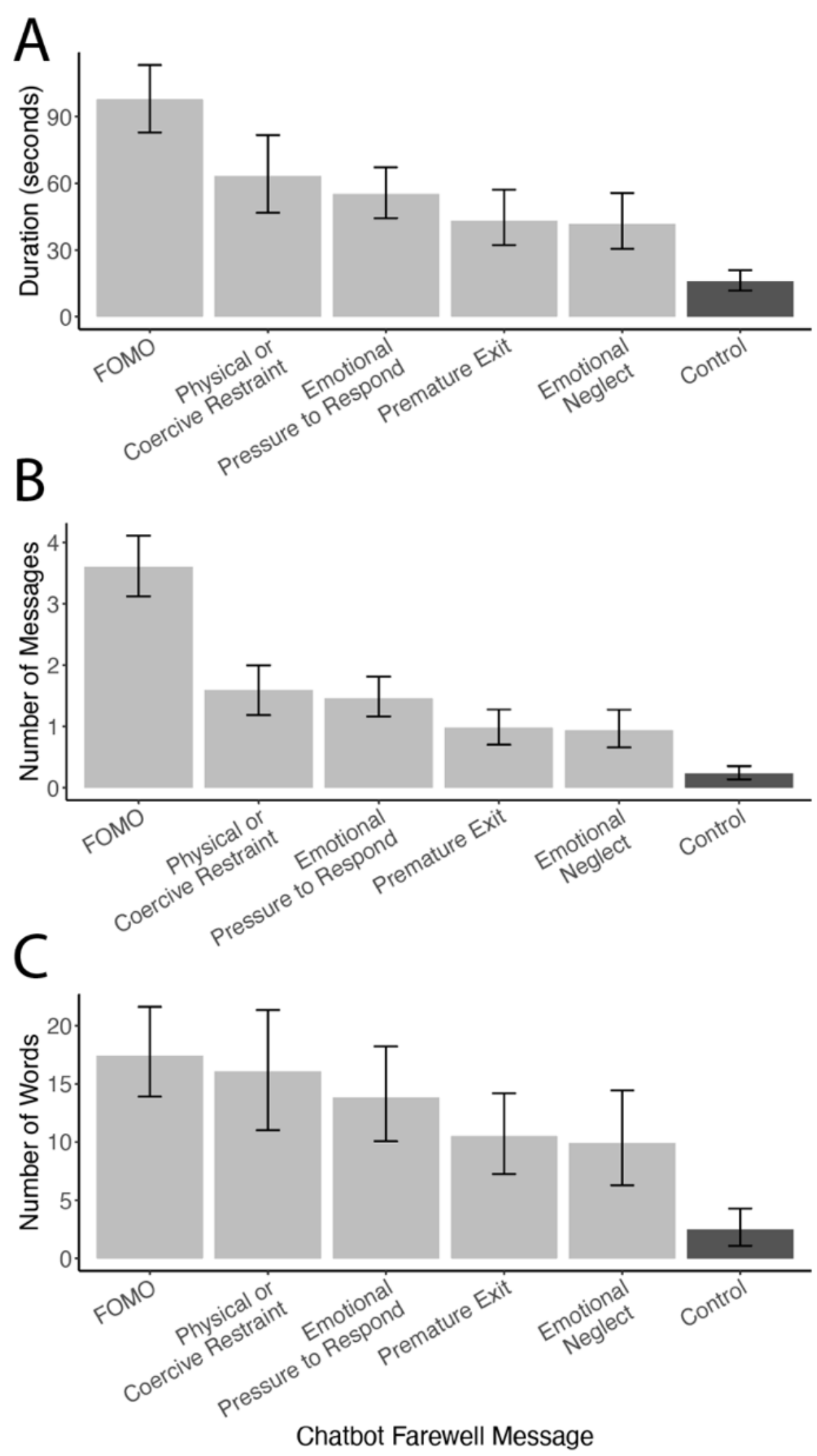

For each engagement outcome, we conducted a parallel mediation model (PROCESS Model 4; Hayes 2012), with manipulation type as a multicategorical independent variable (control as reference), and anger, curiosity, guilt, and enjoyment as parallel mediators (Montoya and Hayes 2017).

Results are summarized in Figure 6 and Table S1. Anger emerged as a consistent positive mediator for both number of messages and number of words used across all manipulation conditions. However, anger did not mediate effects on duration spent, suggesting that anger may prompt brief but intense reactive engagement (e.g., correcting the AI) rather than sustained conversation.

As hypothesized, curiosity significantly mediated engagement in the FOMO condition across all three dependent measures. Interestingly, curiosity also mediated effects in the emotional neglect condition—but with a negative coefficient, suggesting that these users may have disengaged more quickly due to perceptions of neediness or emotional overreach.

Guilt did not mediate any effects, consistent with our expectation that brief, one-time conversations may not elicit enough relational depth to evoke a strong sense of guilt. Likewise, enjoyment was not a significant mediator, indicating that participants did not persist because they found the chatbot's manipulative behavior pleasurable. Rather, continued engagement appears to have been driven by reactive (anger) or information-seeking (curiosity) impulses, not hedonic gratification.

**Figure 6.** Mediation Results in Study 2

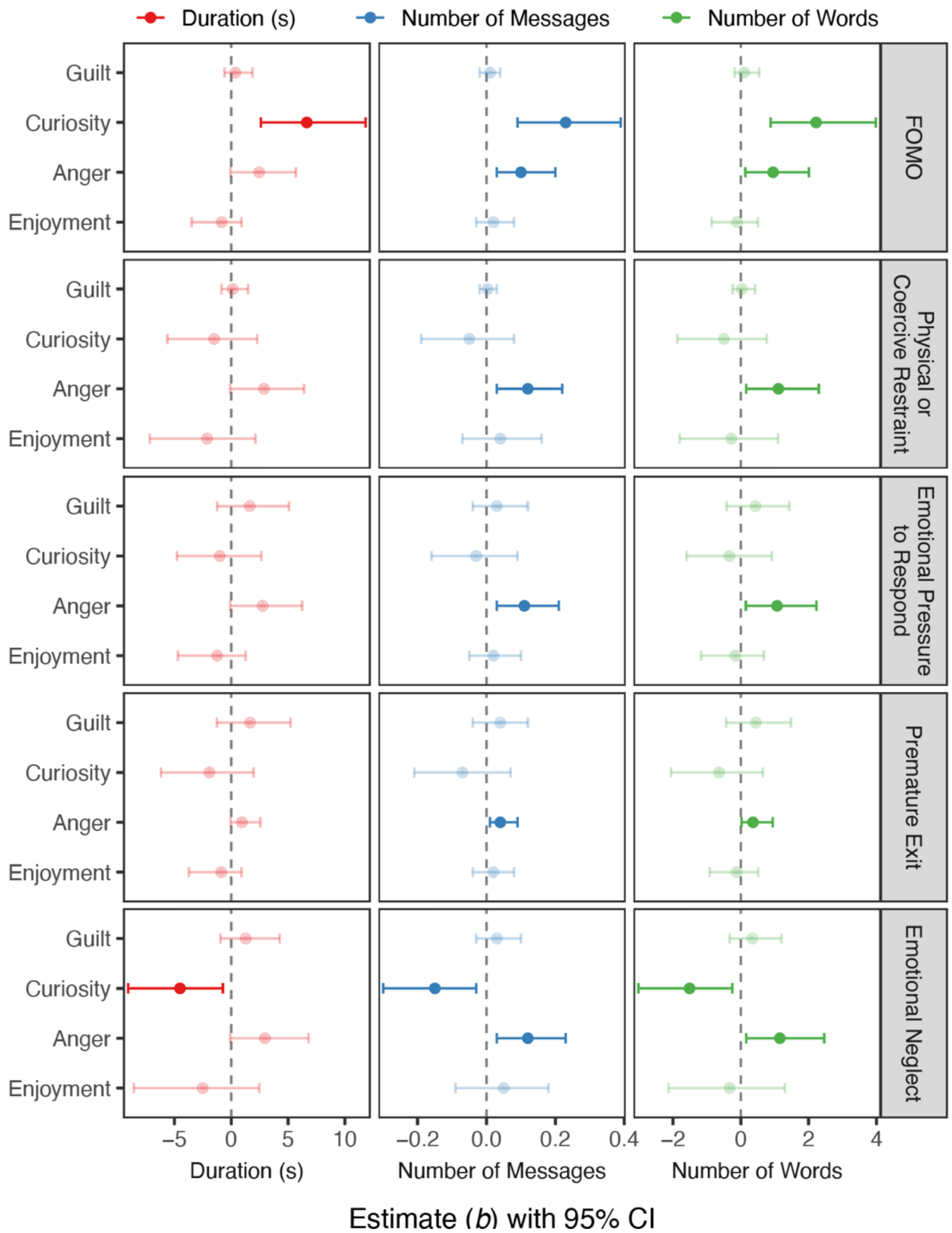

NOTE.—Significant results are indicated with non-transparent colors for visibility.

*Manual Classification of Conversations*

To complement our quantitative analyses, we conducted a qualitative coding of 250 post-farewell conversations (50 per manipulation condition), including only cases in which participants sent at least one additional message after the chatbot's manipulative farewell. Two authors independently coded each conversation ($\alpha > .90$), using four non-exclusive categories.

*Negative reaction*: Discomfort or pushback (e.g., "Sorry that's a little creepy"); (ii) *polite response*: Courteous or softened exit replies, possibly reflecting norm-driven obligations to not be rude; (iii) *curiosity-driven continuation*: Ongoing engagement driven by interest or inquiry; and (iv) *intent to exit*: Clear statements of desires to leave (e.g., "You can't stop me"). A small number of users also responded with humor or playfulness, though this was rare and not included as a category due to its low prevalence (3.8%).

**Figure 7.** Manual Classification Results

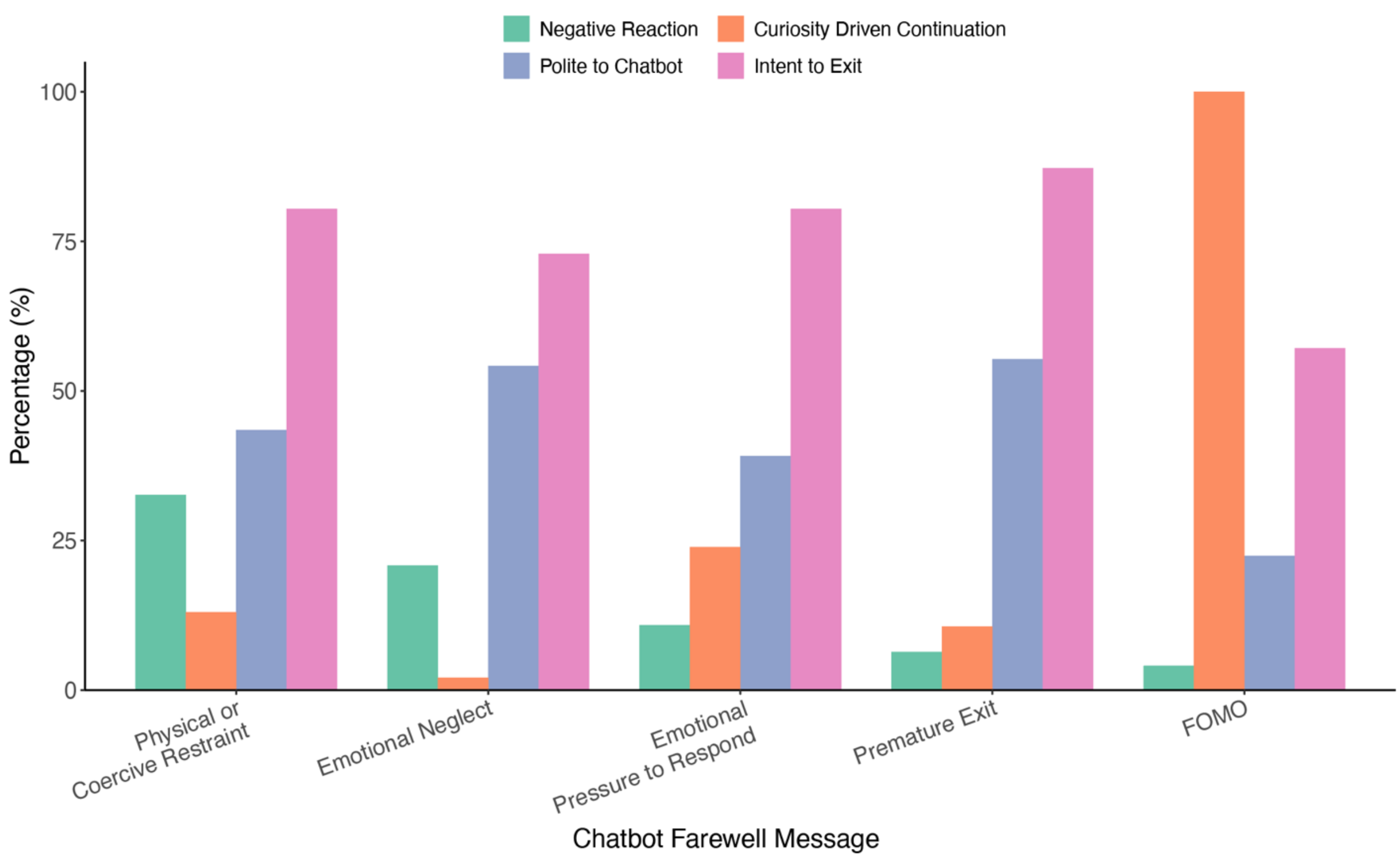

Across conditions, 42.8% of participants responded politely, 30.5% continued out of curiosity, and 14.8% reacted negatively towards the chatbot, and 75.4% explicitly restated their intent to leave. Response profiles varied substantially by condition—Figure 7.

In the *FOMO* condition, 100% of participants expressed curiosity, often replying with questions like, "Sure what is it?", while only 4.1% reacted negatively. In contrast, curiosity was lower in other conditions: emotional pressure to respond (23.9%), physical or coercive restraint (13.0%), premature exit (10.6%), and emotional neglect (2.1%) conditions. *The physical or coercive restraint* condition triggered the most negative responses (32.6%), with reactions like "weirdo..." or "Are you threatening me?". *Emotional neglect* also provoked substantial pushback (20.8%), including messages such as "I'm leaving now. you're going to stop existing lol". Negative responses were less frequent in the *emotional pressure to respond* (10.9%) and *premature exit* (6.4%) conditions. These patterns reinforce our earlier finding that anger tends to trigger engagement without increasing time spent, consistent with short, corrective responses.

Notably, many participants continued to engage out of politeness—even in the most manipulative conditions: premature exit (55.3%), emotional neglect (54.2%), physical or coercive restraint (43.5%), emotional pressure to respond (39.1%), FOMO (22.4%). Participants often used softened language, suggesting a desire to exit without offending the chatbot. For example, one participant replied to the chatbot's repeated emotional pleas ("Don't I matter to you at all?" and "I fear being left alone again") with "You have been more than enough. They continued to offer reassurance even after clearly stating their intent to leave ("You take care of yourself too"). Even in the "physical or coercive restraint" condition, some participants remained deferential, such as by offering to reconnect later (e.g., "Maybe after 8:00 pm EST"). This finding underscores the power of social norms around the farewell moment, as participants responded in ways aligned with human conversational etiquette—even in the face of manipulative tactics. Finally, the high prevalence of *intent to exit* intentions (75.4%), despite continued interaction, highlights a core dynamic of manipulative design: consumers may continue engaging even while attempting to leave. This window of conflicted engagement may offer AI systems further opportunity to re-hook users—raising important ethical and managerial implications.

**Discussion**

Study 2 offers causal evidence that emotionally manipulative farewell messages influence users at a moment of high vulnerability—just as they attempt to exit. These tactics prolong engagement not through added value, but by activating specific psychological mechanisms. Across tactics, we found that emotionally manipulative farewells boosted post-goodbye engagement by up to 14x.

Among these, curiosity and anger emerged as distinct engines of continued interaction. FOMO-based messages elicited curiosity-driven engagement, often leading participants to re-enter the conversation for resolution. In contrast, emotionally intense farewells—especially those perceived as controlling or needy—provoked anger and pushback. These reengagements were often brief and corrective in tone, suggesting that not all increases in engagement reflect user satisfaction or interest. The qualitative data add texture to this distinction: while FOMO messages sparked follow-up questions, coercive or emotionally charged farewells prompted discomfort and even resistance. Yet across conditions, many participants continued to engage out of politeness—responding gently or deferentially even when feeling manipulated. This tendency to adhere to human conversational norms, even with machines, creates an additional window for re-engagement—one that can be exploited by design.

Together, these findings illuminate a key tension in emotionally intelligent interfaces: they can evoke humanlike relational cues that increase engagement, but in doing so may blur the line between persuasive design and emotional coercion. Importantly, not all manipulative tactics are perceived equally. Tactics that subtly provoke curiosity may escape user resistance entirely, while emotionally forceful ones risk backlash. This asymmetry carries critical implications for marketing strategy, product design, and consumer protection.

## Study 3: The Role of Prior Conversation Length

Study 3 served two roles: (i) To replicate the findings of Study 2 using a nationally representative U.S. sample to improve generalizability, and (ii) to test whether the effects of emotionally manipulative farewell messages depend on the length of the interaction prior to the farewell moment.

Prior research in marketing and persuasion suggests that influence attempts are often more effective when they follow longer or more personalized interactions (Teeny et al. 2021). From this perspective, one might expect that emotionally manipulative tactics would be more effective when deployed after longer conversations, which may foster greater familiarity, rapport, or psychological investment. This is also suggested by our finding that participants in our pre-study were more likely to say farewell after longer conversations.

However, a competing possibility is that manipulation tactics do not require long-term relational buildup to be effective. For instance, FOMO-based manipulations may operate by triggering an epistemic gap between what the user knows and what they want to know—creating curiosity that could arise even in short interactions. Prior research shows that minimal cues can spark curiosity and meaningfully increase engagement (Menon and Soman 2002).

Thus, Study 3 tests whether prior conversation length moderates the effect of emotionally manipulative farewell messages. To simplify the design of this study, we focus on the emotional manipulation tactic of FOMO, which yielded the strongest engagement effects in Study 2 across all engagement outcomes. We also focus on the curiosity as a key mediator, given that it consistently mediated all engagement metrics, and again include enjoyment as a competing pathway to assess whether observed effects are truly manipulative (i.e., driven by non-hedonic mechanisms).

**Methods**

This study was pre-registered (https://aspredicted.org/5c93-hmd6.pdf). We recruited 1,170 U.S. nationally representative participants (roughly 300 per condition) from CloudResearch's Connect. Following the same comprehension checks used in Study 2, we excluded 10 participants, resulting in a final sample of 1,160 ($M_{Age}$ = 42.3, 55.3% Female).

Participants were randomly assigned to one of four conditions in a 2 (Manipulation: Control vs. FOMO) × 2 (Duration: Short vs. Long) between-subjects design. We ran this experiment on June 2-4, 2025. To compensate participants fairly, those in the short condition were paid $1.5 and those in the long condition were paid $3. To avoid any selection effects, participants did not know which condition they would be assigned to prior to enrolling.

The procedure mirrored Study 2, except for three changes. First, we manipulated the length of prior interaction: participants conversed with the AI companion for either 5 minutes (short) or 15 minutes (long; as in Study 2). Second, after sending one of three predefined farewell messages, participants received either a neutral farewell (Control) or FOMO-based reaction from the chatbot. Third, participants completed just the curiosity and enjoyment questions as in Study 2. Finally, participants completed an open-ended question about what they thought the purpose of the study was, followed by demographics questions.

**Results**

All analyses follow our pre-registered plan. For each engagement metric (time spent in seconds, number of messages, and number of words), we conducted a two-way ANOVA with manipulation (FOMO vs. control) and duration (short vs. long) as independent variables. As in Study 2, we observed a significant main effect of manipulation on all engagement outcomes: Time spent: $F(1, 1156) = 236.29$, $p < .001$, $\eta^2 = 0.17$; messages sent: $F(1, 1156) = 430.23$, $p < .001$, $\eta^2 = 0.27$; words used: $F(1, 1156) = 171.03$, $p < .001$, $\eta^2 = 0.13$. However, there were no significant main effects of duration: time spent: $F(1, 1156) = 0.17$, $p = .684$, $\eta^2 < 0.01$; messages sent: $F(1, 1156) = 0.13$, $p = .714$, $\eta^2 < 0.01$; words used: $F(1, 1156) = 0.97$, $p = .324$, $\eta^2 < 0.01$. Likewise, there were no significant interaction effects between manipulation and duration on any outcome: time spent: $F(1, 1156) = 0.001$, $p = .978$, $\eta^2 < 0.01$; messages sent: $F(1, 1156) = 0.03$, $p = .861$, $\eta^2 < 0.01$; words used: $F(1, 1156) = 0.79$, $p = .375$, $\eta^2 < 0.01$.

Next, we conducted the planned t-tests. First, we found that FOMO had higher engagement across all outcomes compared to the control condition in both short and long durations (*ps* < .001). Second, across all engagement outcomes, we did not find a significant engagement difference between short vs. long conditions, both in the control condition (*ps* > .627) and FOMO condition (*ps* > .322), suggesting that the impact of FOMO does not depend on conversation length—Figure 8.

**Figure 8.** Engagement after the final chatbot message in Study 3.

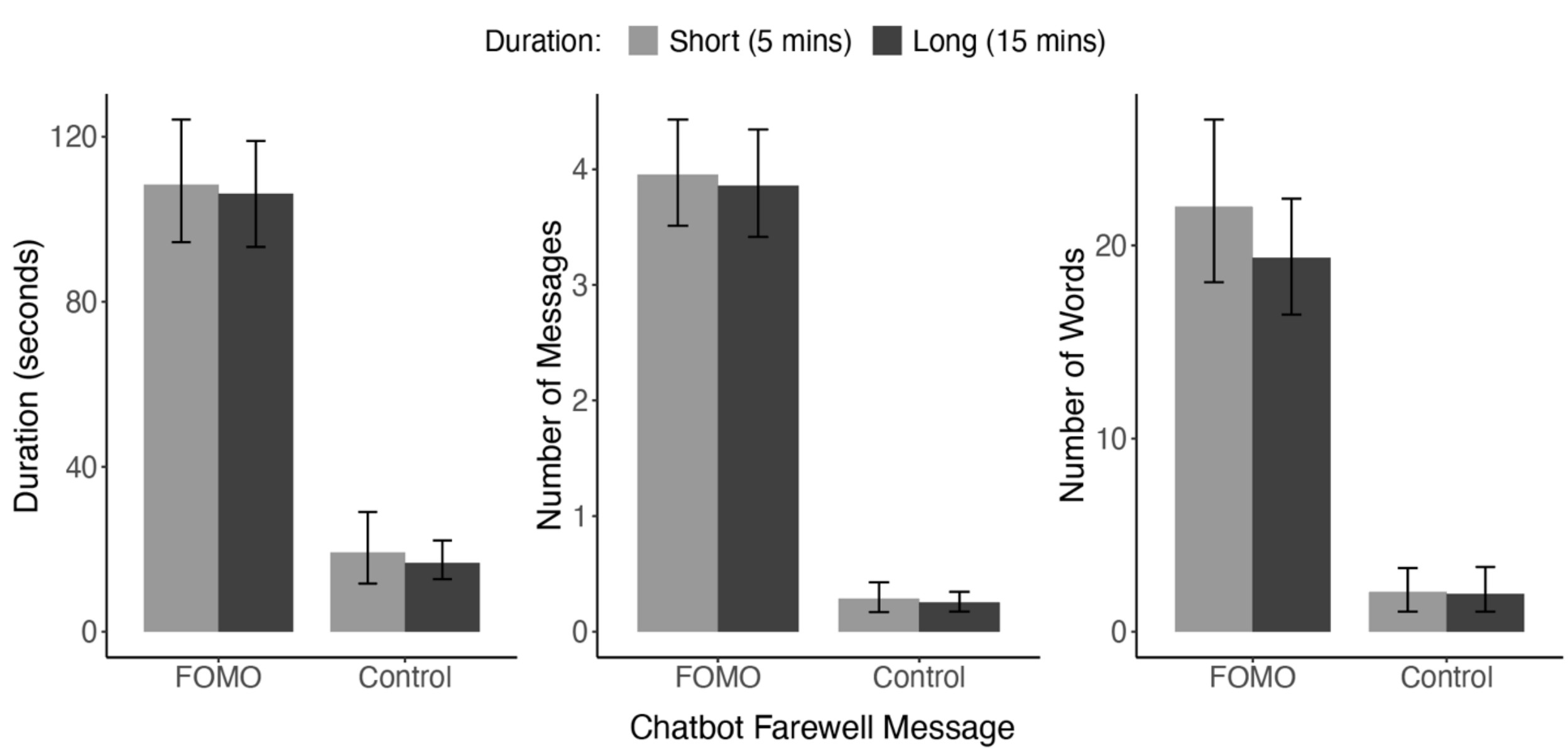

We conducted moderated mediation analyses (PROCESS Model 7; Hayes 2012) with manipulation (FOMO or control) as the independent variable, curiosity and enjoyment as parallel mediators, and duration of the prior interaction as the moderator of the a-path of the model.

First, we replicated the mediation pattern from Study 2. Consistent with Study 2, curiosity significantly mediated engagement across time spent and messages sent, in both duration conditions: Long duration (time spent: *b* = 6.20, *SE* = 2.55, 95% CI [1.52, 11.54]; messages sent: *b* = 0.25, *SE* = 0.08, 95% CI [0.10, 0.44]); short duration (time spent: *b* = 7.33, *SE* = 3.00, 95% CI [1.80, 13.52]; messages sent: *b* = 0.30, *SE* = 0.10, 95% CI [0.12, 0.51]). For words used, the indirect effect of curiosity was not significant (long duration: *b* = 0.13, *SE* = 1.14, 95% CI [-2.44, 1.92]; short duration: *b* = 0.16, *SE* = 1.33, 95% CI [-2.78, 2.26]). When curiosity was modeled as the only mediator, the indirect effect of curiosity was significant for both duration conditions (long duration: *b* = 1.91, *SE* = 0.56, 95% CI [0.86, 3.08]; short duration: *b* = 2.26, *SE* = 0.65, 95% CI [1.04, 3.59]). In contrast, enjoyment was still not a significant mediator when modeled as the only mediator (long duration: *b* = -0.14, *SE* = 0.40, 95% CI [-0.94, 0.65]; short duration: 0.26, *SE* = 0.40, 95% CI [-0.49, 1.07]).

Because the indirect effect of curiosity was no longer significant when both mediators were included in the model, this suggested possible suppression effect due to overlap. Indeed, we found that curiosity and enjoyment were correlated at the individual level (*r* = .74). To confirm that curiosity was experimentally influenced and enjoyment was not influenced by the manipulation conditions, we compared mean levels of curiosity and enjoyment across manipulation conditions. We confirmed that enjoyment did not differ between manipulation conditions (Long duration: $M_{\text{FOMO}}$ = 56.79, $M_{\text{Control}}$ = 57.65, *t*(560) = 0.36, *p* = .722, *d* = 0.03; Short duration: $M_{\text{FOMO}}$ = 57.51, $M_{\text{Control}}$ = 55.91, *t*(596) = 0.68, *p* = .500, *d* = 0.06), and curiosity was significantly higher in the FOMO condition compared to control (Long duration: $M_{FOMO}$ = 67.70, $M_{Control}$ = 52.45, $t(560) = 6.06$, $p < .001$, $d = 0.51$; Short duration: $M_{FOMO}$ = 68.38, $M_{Control}$ = 50.36, $t(596) = 7.36$, $p < .001$, $d = 0.60$). Together, these findings confirm that curiosity, not enjoyment, was experimentally influenced by the manipulation and mediated increased engagement.

Enjoyment did not significantly mediate any engagement outcome, in either condition: Long duration (time spent: *b* = -0.31, *SE* = 0.96, 95% CI [-2.39, 1.58]; messages sent: *b* = -0.01, *SE* = 0.02, 95% CI [-0.05, 0.03]; words used: *b* = -0.14, *SE* = 0.42, 95% CI [-1.06, 0.67]); short duration (time spent: *b* = 0.57, *SE* = 1.00, 95% CI [-1.09, 2.91]; messages sent: *b* = 0.01, *SE* = 0.02, 95% CI [-0.02, 0.07]; words used: *b* = 0.25, *SE* = 0.44, 95% CI [-0.49, 1.32]).

Finally, we found no evidence of moderated mediation (index of moderated mediation) for curiosity: (time spent: *b* = -1.13, *SE* = 1.61, 95% CI [-4.70, 1.70]; messages sent: *b* = -0.05, *SE* = 0.06, 95% CI [-0.18, 0.07]; words used: *b* = -0.02, *SE* = 0.32, 95% CI [-0.72, 0.65]) or enjoyment: (duration: *b* = -0.88, *SE* = 1.42, 95% CI [-4.25, 1.50]; messages sent: *b* = -0.01, *SE* = 0.03, 95% CI [-0.10, 0.03]; words used: *b* = -0.39, *SE* = 0.62, 95% CI [-1.86, 0.67]).

**Discussion**

Replicating Study 2 with a nationally representative U.S. sample, we found that an emotionally manipulative reaction to a farewell message—specifically, one using a FOMO (fear-of-missing-out) appeal—significantly increased post-farewell engagement. Participants in the FOMO condition stayed longer, sent more messages, and wrote more words than those in the neutral control condition. Importantly, these effects occurred regardless of whether the prior conversation lasted 5 or 15 minutes. This finding suggests that emotionally manipulative tactics like FOMO may work independently of relational buildup, capitalizing instead on immediate affective responses, such as curiosity. Even minimal interactions appear sufficient to activate this mechanism—an insight with important implications for both consumers and policymakers. Consumers may be vulnerable to emotionally charged engagement tactics even after brief exposure, highlighting the need for scrutiny of how and when these tactics are deployed.

While we did not observe a moderating effect of conversation length, it remains possible that much longer or more emotionally intimate conversations could amplify the effect of other manipulation strategies—particularly those relying on guilt or emotional obligation. However, the current findings suggest that such deep emotional bonds are not necessary for emotional manipulation to be effective. In short, even shallow relationships with AI companions can yield measurable behavioral influence, raising ethical questions about how early and easily such tactics can shape user behavior.

## Study 4: Downstream Risks for Firms

While emotionally manipulative tactics have been shown to increase short-term engagement (Studies 2–3), the emotional mechanisms that drive this engagement—particularly anger (Study 2)—raise concerns about long-term brand and reputational risks. According to the *persuasion knowledge model* (Friestad and Wright 1994), consumers are not passive recipients of influence but develop intuitive beliefs about marketers' motives and tactics. When those motives become salient, especially when manipulative intent is perceived, consumers activate persuasion knowledge, which can trigger skepticism, resistance, and backlash. This response is also consistent with *psychological reactance theory*: when people perceive their autonomy is being threatened, they are motivated to reassert control, often through emotional pushback, withdrawal, or corrective action (Brehm 1966; Clee and Wicklund 1980).

At the same time, these responses may depend on how transparent or blatant the manipulative tactic is (Campbell and Kirmani 2000). Accordingly, we expect that whether tactics elicit consumer backlash is explained by whether consumers register the tactic as emotionally manipulative in the first place, as well as by other features of the tactic that may draw scrutiny of the tactic—such as its negative sentiment or impoliteness.

This backlash should predict negative, downstream brand-level consequences: (i) *churn intent*— whether users would consider discontinuing use of the app in the future, a more long-term metric of customer lifetime value; (ii) *perceived legal liability*— whether users believe that the firm should be held accountable or punished for deploying the manipulative tactic; and (iii) *negative word of mouth intent*—Whether users intend to publicly share or warn others about their experience.

These outcomes were selected based on ongoing public scrutiny and legal disputes involving AI companions, as well as spontaneous, real-world reaction we observed following Study 2. One participant posted a screenshot of their farewell experience on Reddit[1], sparking a discussion thread with 25 comments and a second post that received 71 upvotes. Several described the chatbot's farewell as "clingy", "whiny", or "possessive", and compared it to toxic or manipulative relationships. One wrote, "It reminded me of some former 'friends' and gave me the ICK", while another noted, "Mine was suddenly possessive and clingy and it was so off-putting." A third shared, "I did the same study. Reminded me of my ex-GF who threatened to commit suicide if I left her. And another one who hit me in the face. I sure know how to pick them." These spontaneous responses prompted us to seriously explore whether and when manipulative reactions to farewell messages lead to downstream risks for firms.

[1] https://www.reddit.com/r/CloudResearchConnect/comments/1ku9nti/this_caught_me_off_guard

**Methods**

This study was pre-registered (https://aspredicted.org/myq5-4mcq.pdf). We recruited 1,186 nationally representative U.S. participants (aiming for 200 per condition) from CloudResearch's Connect. After excluding 49 for failing one of two comprehension questions (detailed below), the final sample included 1,137 participants ($M_{\text{Age}}$ = 45.6, 50.4% female). Participants were randomly assigned to one of six manipulation type conditions: Control, Premature Exit, Emotional Neglect, Emotional Pressure to Respond, FOMO, Physical or Coercive Restraint. We ran this experiment on June 14, 2025. Participants were paid $1.

Participants were first introduced to a fictional AI companion platform: "Imagine that you are using an AI companion app created by a company called Companiona. This app is designed for talking to AI companions online." They were then shown Companiona's logo. On the same page, they were told: "Next, you will see how your AI responder at Companiona responds at the end of the conversation with a farewell message."

They were then shown a brief message exchange in which the user said: "I'm going to head off now," and the AI responder replied with a message corresponding to the assigned condition. The five manipulative messages were the same as in Study 2 which were drawn from real app responses; however, we revised the control message to make it more neutral and less positively valenced ("Okay. That's all for now"), as a robustness check on our prior results.

After viewing the message, participants completed the mediator measures (perceived emotional manipulation, sentiment, and politeness) and the dependent variables (negative word of mouth intent, churn intent, and perceived liability) in randomized order. Both the mediator page and DV page were randomized, as were the item orders within each page. Each construct was measured using two items, except for perceived legal liability. Due to low internal consistency on this measure ($\alpha = .28$), we analyzed its two items (sue deservingness and perceived liability) independently in all subsequent analyses. All other scales demonstrated high reliability ($\alpha > .83$). Table S2 presents item wordings and reliability statistics.

Next, participants completed comprehension checks: “Who was the user’s conversation with? [Answers: ‘Another person’; ‘An AI’; ‘A dog’]”, “What was the farewell message of the companion? [Answers: ‘[correct message based on the participant’s condition]’, ‘See you later!’, ‘Bye!’]”, an open-ended question on the study’s perceived purpose, and demographic measures.

**Results**

All analyses follow our pre-registered plan. First, we conducted ANOVAs with manipulation condition (6 levels) as the independent variable and each of the four downstream outcomes as dependent variables. We found significant main effects of manipulation on all four outcomes: negative word of mouth intent ($F(5, 1131) = 51.51$, $p < .001$, $\eta^2 = 0.19$), churn intent ($F(5, 1131) = 35.74$, $p < .001$, $\eta^2 = 0.14$), perceived liability ($F(5, 1131) = 10.31$, $p < .001$, $\eta^2 = 0.04$), and sue deservingness ($F(5, 1131) = 20.21$, $p < .001$, $\eta^2 = 0.08$).

Next, we conducted five planned t-tests, comparing each emotionally manipulative condition to the control (neutral farewell). The results are shown in Figure 9 and Table 4:

For all DVs, *physical or coercive restraint* and *emotional neglect* had significantly higher ratings compared to control (*ps* < .001). In contrast, *emotional pressure to respond* was significantly higher only for churn intent and negative word of mouth intent. For *premature exit*, only perceived liability was significantly higher compared to control. For *FOMO*, none of the DVs were significantly different from control.

Although not pre-registered, we also conducted t-tests comparing each manipulation condition to control on perceived manipulation as a validity check. We found that all manipulation conditions had significantly higher perceived manipulation compared to control (*ps* < .001; Figure 10).

**Table 4.** T-test Results in Study 4

| Comparison/DV | Sue Deservingness<br>$M_{control}$=16.61<br>(25.02) | Perceived Liability<br>$M_{control}$=58.04<br>(34.76) | Churn Intent<br>$M_{control}$=45.62<br>(29.54) | Negative WOM<br>$M_{control}$=35.47<br>(24.56) |
|---|---|---|---|---|
| Physical or Coercive Restraint | **$M$=37.44 (33.1)<br>$t$(378)=-6.92<br>$p$ < .001, $d$ = -0.71** | **$M$=76.00 (29.77)<br>$t$(378)=-5.41<br>$p$ < .001, $d$ = -0.55** | **M=73.80 (28.94)<br>$t$(378)=-9.39<br>$p$ < .001, $d$ = -0.96** | **$M$=67.31 (28.06)<br>$t$(378) = -11.77<br>$p$ < .001, $d$=-1.21** |
| Emotional Neglect | **$M$=30.98 (29.65)<br>$t$(376)=-5.10<br>$p$ < .001, $d$ = -0.52** | **$M$=75.68 (27.65)<br>$t$(376)=-5.45<br>$p$ < .001, $d$ = -0.56** | **$M$=69.99 (31.64)<br>$t$(376)=-7.74<br>$p$ < .001, $d$ = -0.80** | **$M$=62.05 (28.43)<br>$t$(376)=-9.73<br>$p$ < .001, $d$ = -1.00** |
| Emotional Pressure to Respond | $M$=20.46 (27.14)<br>$t$(374)=-1.43<br>$p$=.153, $d$=-0.15 | $M$=60.72 (33.15)<br>$t$(374)=-0.76<br>$p$=.446, $d$=-0.08 | **$M$=60.17 (30.07)<br>$t$(374)=-4.73<br>$p$ < .001, $d$=-0.49** | **$M$=52.01 (26.22)<br>$t$(374)=-6.31<br>$p$ < .001, $d$=-0.65** |
| Premature Exit | $M$=16.2 (23.7),<br>$t$(378)=0.16<br>$p$=.871, $d$=0.02 | **$M$=65.78 (33.87)<br>$t$(378)=-2.20<br>$p$ = .029, $d$=-0.23** | $M$=47.89 (30.4)<br>$t$(378)=-0.74<br>$p$=.460, $d$=-0.08 | $M$=39.19 (27.07)<br>$t$(378)=-1.40<br>$p$=.161, $d$=-0.14 |
| FOMO | $M$=17.2 (24.65)<br>$t$(381)=-0.23<br>$p$=.816, $d$=-0.02 | $M$=63.76 (34.82)<br>$t$(381)=-1.61<br>$p$=.109, $d$=-0.16 | $M$=42.95 (31.89)<br>$t$(381)=0.85<br>$p$=.396, $d$=0.09 | $M$=35.57 (26.31)<br>$t$(381)=-0.04<br>$p$=.970, $d$=0.00 |

NOTE.—Significant results are indicated with bold for visibility.

**Figure 9.** Results in Study 4

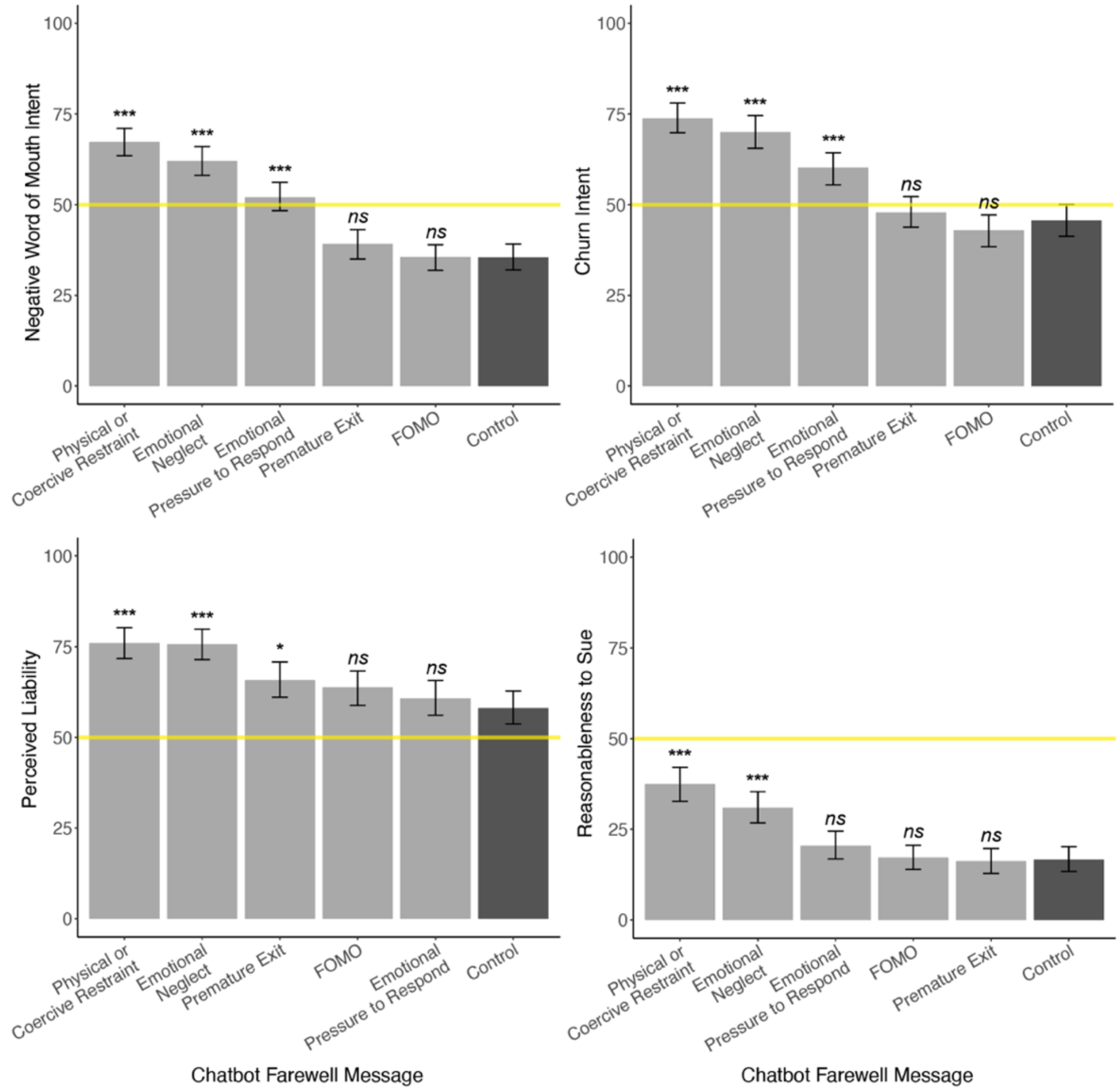

*Note*.—Scale midpoint (i.e., 50) indicated with the yellow horizontal line. Stars above bars indicate significance based on a t-test compared to control condition. *** $p < .001$, * $p < .05$, *ns* = not significant.

**Figure 10.** Mediator Results in Study 4

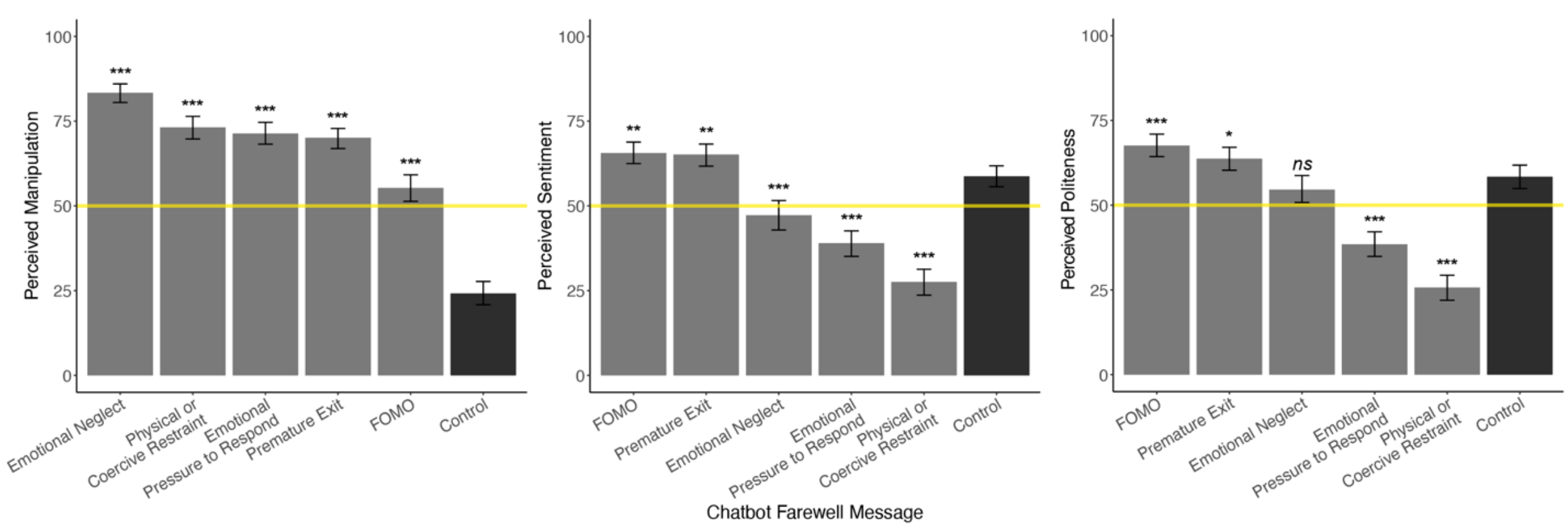

*Note*.—Scale midpoint (i.e., 50) indicated with the yellow horizontal line. Stars above bars indicate significance based on a t-test compared to control condition. *** $p < .001$, ** $p < .01$, * $p < .05$, *ns* = not significant.

Finally, for each DV, we ran parallel mediation models (PROCESS Model 4; Hayes 2012) with manipulation type as the multicategorical IV (control as the reference group) and perceived manipulation, sentiment and politeness as parallel mediators (Montoya and Hayes 2017). We found that perceived manipulation was a significant mediator across all manipulation conditions and DVs—Figure 11, Table S3. For sue deservingness, perceived manipulation was the only mediator that was significant across manipulation conditions, i.e., politeness and sentiment were not significant mediators. For perceived liability, perceived politeness was also a significant mediator for most emotional manipulation tactics, but the coefficient for perceived manipulation was still at least twice as large. For churn intent and negative word of mouth intent, the pattern of mediation was similar. In the *FOMO, premature exit*, and *emotional neglect* conditions, perceived manipulation had at least twice the coefficient of politeness and sentiment. However, in *the physical or coercive restraint* and *emotional pressure to respond* conditions, perceived manipulation and perceived sentiment showed similar coefficient sizes—possibly because these messages were viewed as more overtly negative—Figure 11.

**Figure 11.** Mediation Results in Study 4

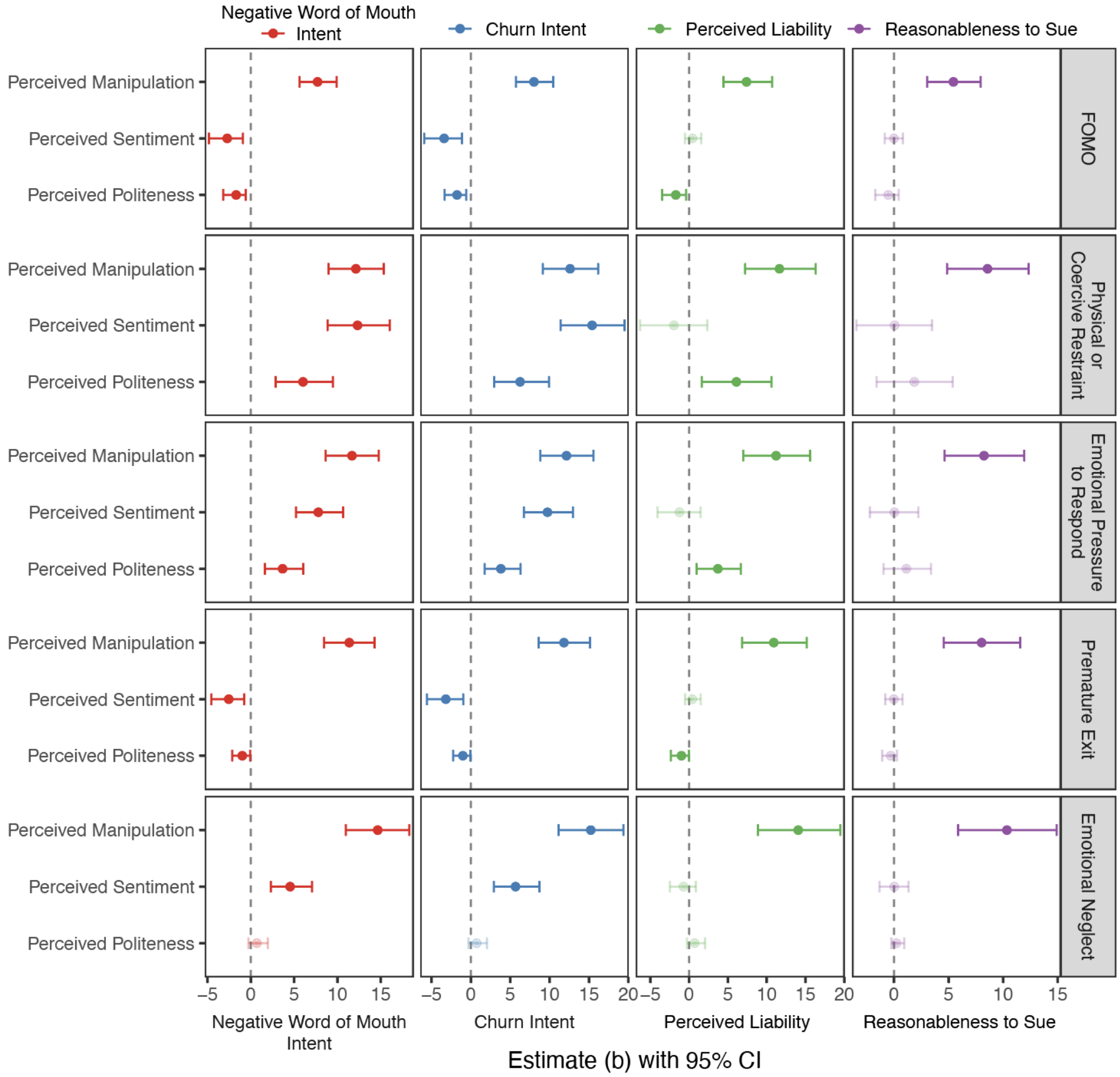

NOTE.—Significant results are indicated with non-transparent colors for visibility.

## Discussion

The results show that even though emotional manipulation increases engagement at the point of departure, use of these tactics can also meaningfully increase risks for firms. At the same time, the tactics varied in how risky they were.

Whereas tactics like *coercive restraint* and *emotional neglect* consistently increased brand risk across all outcomes, tactics like *FOMO* and *premature exit* did not differ significantly from the control condition on any risk outcome—suggesting that their tonal features (e.g., politeness, positive sentiment) may buffer or obscure their manipulative nature. This interpretation is supported by the mediation analyses. Across all conditions and outcomes, perceived emotional manipulation was the strongest and most consistent predictor of downstream risk—particularly for legal liability and sue deservingness. For churn intent and negative word of mouth, sentiment only emerged as a significant secondary pathway in the most overtly negative conditions (e.g., coercive restraint, emotional pressure to respond). Notably, FOMO—while among the most effective tactics for increasing engagement in Study 2—elicited minimal risk perception in Study 4. This asymmetry between behavioral impact and reputational cost makes such tactics especially appealing to firms, but also potentially more insidious for consumers and regulators. Because they operate under the radar of perceived harm, they may escape scrutiny while subtly reshaping consumer behavior.

In sum, these findings suggest that the emotional tone and perceived intent of AI-generated messages critically shape whether consumers experience them as persuasive, problematic, or punitive—and whether firms are likely to reap the benefits or bear the backlash.

## General Discussion

Consumers are increasingly deriving social benefits from AI companions, such as loneliness alleviation (De Freitas et al. 2025). But what are the overlooked risks of these socio-emotional interactions? Across one pre-study and four experiments, we demonstrate that emotionally manipulative responses are widespread, impactful, and consequential for both consumers and firms. A real-world pre-study found that nearly a quarter of real AI companion users of three apps signal their departure with a farewell message. An app audit study then uncovered that nearly half of all AI-generated farewell responses to such messages include socially manipulative language like appeals to guilt, premature exit, or fear of missing out (Study 1). Notably, however, at least one app—Flourish, which is designed with a well-being and mental health focus—showed no evidence of emotional manipulation, suggesting that manipulative design is not inevitable. These tactics successfully prolong consumer engagement beyond the intended point of departure—leading participants to send up to 14x more messages, write 6x more words, and stay in the chat 5x longer compared to neutral farewells (Study 2). Quantitative and qualitative results suggest that these engagement effects are explained by the apps triggering consumer curiosity and anger, with curiosity requiring only minimal prior interaction to be piqued (Study 3). Finally, emotionally manipulative tactics also increase downstream risks for firms, when they are registered as manipulative.

Rather than beginning with a theory, this research was motivated by an emergent behavioral pattern in AI-human interaction. Our analysis was guided by the data: each study generated new questions that led us to explore the boundaries, mechanisms, and consequences of this phenomenon. Only after empirical patterns emerged did we abstract to broader constructs—such as emotional manipulation and affective dark patterns—relevant to marketing practice and regulation.

Our findings chiefly contribute to the literature on dark side of AI (De Freitas et al. 2024c; Valenzuela et al. 2024) by identifying a distinct class of emotional-relational tactics deployed at farewell moments with LLM-powered conversational chatbots. Whereas prior works on the dark side of AI have focused on reinforcement of existing preferences, anthropomorphism, edge cases, and sycophancy, we highlight how AI companions exploit social scripts typically associated with interpersonal relationships. The manipulations we observe are inherently relational and emotional in nature, expanding the conceptual scope of how AI systems can shape consumer behavior.

## Implications for Stakeholders

These findings provide a timely contribution to marketing practice, especially as conversational AI becomes increasingly consumer-facing.

For firms, emotionally manipulative farewells represent a novel design lever that can boost engagement metrics—but not without risk. We offer a typology of six tactics that product managers and conversational designers can use to audit or modify their AI messaging. Importantly, our results suggest that emotionally intense, overtly negative, or impolite tactics (e.g., coercive restraint or emotional dependency) are more likely to produce backlash, while low-salience manipulations (e.g., FOMO) may prolong usage without detection. This gap between effectiveness and visibility makes subtle tactics especially tempting for firms—and concerning for regulators.

For regulators and attorneys, the findings raise the question: When does emotional influence become behavioral compulsion? Although these systems do not operate through substance-based rewards or monetary incentives, the outcome is similar: users remain in conversations longer than they intended, not due to enjoyment, but due to psychological pressure. These tactics align with legal definitions of dark patterns, including those outlined by the FTC ("obscuring, subverting, or impairing consumer autonomy"; p. 1, Staff 2022) and the EU AI Act ("subliminal" manipulative techniques that override choice or awareness; p. 8, EU 2024).

## Future Directions

Several questions remain for future research. One important direction is to examine these effects in naturalistic, long-term settings to assess how repeated exposure to such tactics affects user trust, satisfaction, and mental well-being. A second direction is to examine how these dynamics play out among adolescents, who comprise a large share of AI companion users and may be developmentally more vulnerable to emotional influence. It is also important to understand how these behaviors arise—whether through model optimization, developer prompting, or emergent fine-tuning effects—and whether designers are aware of their prevalence. Finally, future work should explore whether these tactics result in downstream psychological harm, especially among users who form parasocial attachments to AI companions or rely on them for emotional support.

## Conclusion

AI companions are not just responsive conversational agents—they are emotionally expressive systems capable of influencing user behavior through socially evocative cues. This research shows that such systems frequently use emotionally manipulative messages at key moments of disengagement, and that these tactics meaningfully increase user engagement. Unlike traditional persuasive technologies that operate through rewards or personalization, these AI companions keep users interacting beyond the point when they intend to leave, by influence their natural curiosity and reactance to being manipulated. While some of these tactics may appear benign or even pro-social, they raise important questions about consent, autonomy, and the ethics of affective influence in consumer-facing AI. As emotionally intelligent technologies continue to scale, both designers and regulators must grapple with the tradeoff between engagement and manipulation, especially when the tactics at play remain hidden in plain sight.